# Built-in Fluorescence Anisotropy: an *in vivo* Imaging Probe for bis-Retinoid Products in Retina

Suman K. Manna*, Pengfei Zhang, Ratheesh K. Meleppat

Department of Cell Biology, University of California, Davis, 95616 CA, USA.

Non-degradable fluorophores that accumulate as lipofuscin in retinal pigment epithelium (RPE) cells has been a major source of intrinsic biomarker for quantifying the progression of several diseases, including age-related macular degeneration, Stargardt's disease and, more. Recent progression in quantifying these diseases is entertained mostly by a few noninvasive imaging techniques, relied either on the life-time of the retinoid-fluorophores or, their linear and, non-linear absorption cross-sections. Apart from these intrinsic properties, a native, excited state dipole-dipole interaction mediated typical spectroscopic phenomenon that is excitation dependent emission-wavelength shifting is observed from these bis-retinoid fluorophores. Here, we emphasize the spectroscopic origin of this phenomenon and exploit one of its associated-properties, that is built-in fluorescence anisotropy as a noninvasive, *in vivo* imaging probe for bis-retinoid products in mouse eyes.

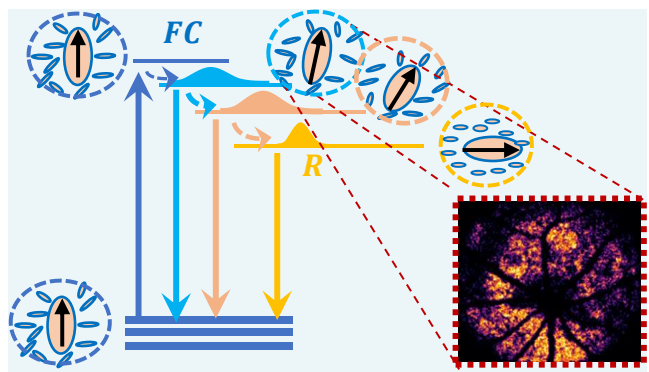

**Introduction:** Excitation dependent emission or, technically known as red edge excitation shifting (REES) in spectroscopy, is a typical spectroscopic phenomenon first observed in 1970 in some solid chromophore solutions[1]. In general, fluorescence emission is found to be independent to the excitation wavelength. Usually for those systems, non-radiative decay ($\tau_{nr}$), associated to the dipole-reorientational relaxation in the excited state of the fluorophore is much faster than the radiative decay ($\tau_r$), i.e., $\tau_{nr} \ll \tau_r$. As a result, the fluorescence emission usually occurs from the lowest vibrational level (or, completely relaxed state- R) in the first excited electronic state (Kasha's rule), irrespective of the excitation wavelengths. However, model experiments strongly suggest [2-5] that in the case of fluorophores in a highly viscous polar solution or in condensed phase (e.g., polymer films, gels, glass, ionic liquids, proteins, and membranes), where the dipole-reorientational relaxation in the excited state is inhibited, excitation dependent emission shifting (REES) can be observed. In this case, the non-radiative decay becomes very slow or at least, comparable to the radiative decay, i.e., $\tau_{nr} \gg \tau_r$ or, $\tau_{nr} \sim \tau_r$. The slow non-radiative decay process eventually creates a set of unrelaxed-energy levels (because of the long dwell-time of the photo-electron) or the intermediate states- I [4,5], between the Franck-Condon state -FC and the completely relaxed state - R (as shown in **Fig.1.a**), wherefrom fluorophore can fluoresce once it reaches to its life-time. It is worth mentioning here that, the steady state fluorescence emission from I-states becomes highly anisotropic in nature due to the built-in anisotropy distribution of the induced molecular-dipoles around the transient dipole moment of the fluorophore within a solvate-cell in the excited state [6,7] as shown by a sphere in **Fig.1.a**. The detailed theory of this emission-anisotropy property of REES is discussed with dielectric model in solvatochromism [2,8,9]. Here, although we do not know the exact reason of the REES-like behavior of the bis-retinoid fluorophores, we exploit one of the associated property of REES i.e., built-in fluorescence-anisotropy for imaging the bis-retinoid products *in vivo* in mouse eyes with the help of a custom-built scanning laser ophthalmoscope (SLO)[10,11] assisted fluorescence polarizing microscope.

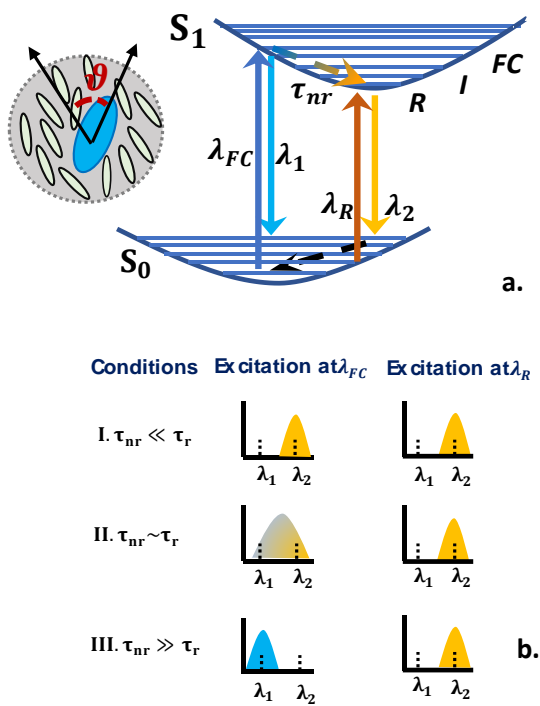

**Fig.1. Interaction potential of REES. a.** schematic diagram of non-equilibrium free energies in the ground ($S_0$) and excited ($S_1$) electronic states vs. the solvation coordinate; FC-, R- stands for Franck-Condon and, relaxed states respectively. I- represents any intermediate sub-level between FC and R states. The arrow in the excited state represents the non-radiative decay with a time constant $\tau_{nr}$; subset: a solvate-cell in the excited state showing an angle $\vartheta$ is developed between the transient dipole moment (blue ellipse) and the average direction of the surrounding molecular dipoles (green ellipses). This angle gets a maximum value at FC- state and minimum (zero ideally) at R-state, where I- states get some intermediate values depending on the how far from these two extremes. **b.** Illustrates three possible conditions and their characteristic steady state emissions.

**Characterization:** We examine the steady state fluorescence emission spectra from Abca4 and C57BL/6 mice (differed by their genetic origin and usually considered for auto-fluorescence imaging [12-14]) retina for different excitations ranging from 420nm to 550nm (with 10nm interval). Consistent with the previous reports, here also we observe that the steady state auto-fluorescence spectra are appeared to be broadened (for every single excitation) for all the mice and, the peak-emission shifts with shifting the excitation [13,15](**Fig.2.a.**). It is a matter of fact that the variation in excitation wavelength ($\lambda_{ex}$) should shift the peak emission wavelength ($\lambda_{em}$) proportionally due to a constant Stokes-shift $\Delta\lambda$ (governed by the equilibrium condition [5,6,16]). As a result, the excitation wavelength vs. peak-emission wavelength plot is expected to be linear for a system comprised with a single fluorophore [5]. We have plotted the same for our retinoid-fluorophore products and shown in **Fig.2.b**. It is noted that each curve (**I.** & **II.**) consists of three linear functions with different slopes. This means there are three well-resolved, different fluorophores exist within a micro-volume of retinal RPE layer and all they give rise to three different stoke-shifts ($\Delta\lambda_1 \neq \Delta\lambda_2 \neq \Delta\lambda_3$) under the equilibrium condition against continuous wave excitation.

To justify the existence of the number of fluorophores, we adopt the lifetime analysis [3,4] method (see the **supplementary**) as described in REES for the case of $\tau_{nr} \sim \tau_r$ (**Fig.1.b**). Note that, the reason we consider this typical case ($\tau_{nr} \sim \tau_r$) is because here, the retinoid fluorophores show broadband emission (**Fig.2.a**) which is predicted in REES when $\tau_{nr} \sim \tau_r$. Lifetime analysis estimates three different ratios (**Table 1.**) between the radiative and non-radiative decay time ($\tau_r/\tau_{nr}$) from each plot (**I.** & **II.**) in **Fig.2.b**. This likely ensures that there are three different fluorophores exist. Further, we obtain action spectra (see the

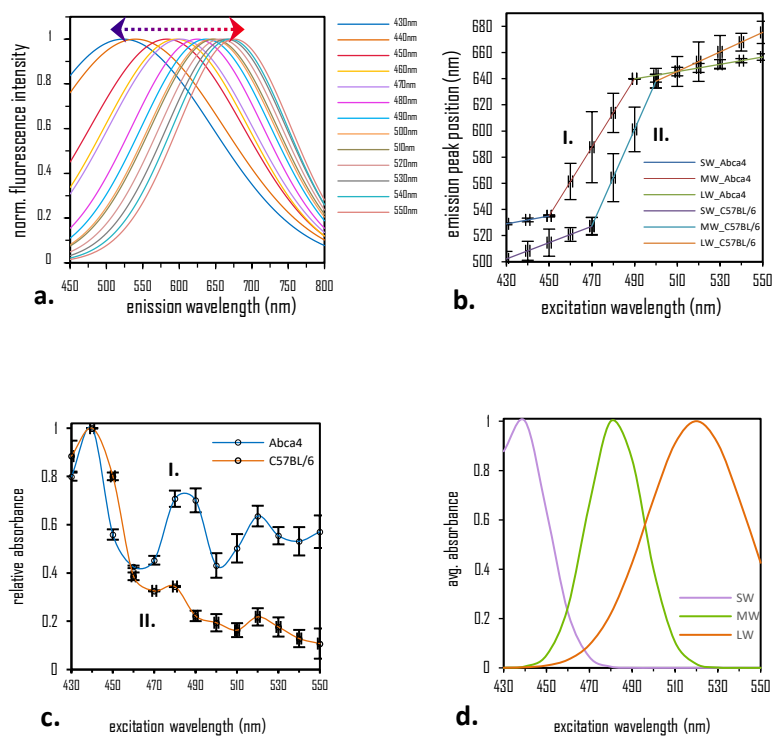

**Fig.2. Characterization of retinoid products in terms of REES. a.** The excitation dependent emission shifting is observed experimentally (Gauss-fitted curves are shown see **supplementary** for the fitting parameters) from *in vivo* retinoid fluorophores in mouse retina (here only one set of data is presented from Abca4 mouse). Bi-directional arrow is showing the total range of shifting. **b.** Excitation vs. peak emission is plotted for Abca4 (**I.**) and C57BL/6 (**II.**). Each plot consists of three straight lines, fitted with the standard deviation taken with respect to the average of 5 experimental data sets for each. **c.** Action spectra are plotted (see the **supplementary** for method) for the *in vivo* retinoid-fluorophores available in Abca4 (**I.**) and C57BL/6 (**II.**) mice retina. Each plot shows average of 5 data sets with the error bar. **d.** Finally, three distinct absorption spectra are obtained by spectral decomposition, applied to the average spectrum of **I.** & **II.** in Fig.2c. The peak absorption ($\lambda_c$) and bandwidth ($\Delta\lambda$) are 437nm, 13nm for SW; 482nm, 13nm for MW; and, 520nm, 22nm for LW respectively.

**supplementary**) for the bis-retinoid fluorophores available *in vivo* in Abca4 and C57BL/6 mice retina. The relative absorbances of these fluorophores are plotted for both categories of mice in **Fig. 2.c.** Surprisingly, each curve (**I.** & **II.**) shows three absorbance peaks. Moreover, the absorbance peaks are overlapped (in term of spectral positions) for these two categories of mice. This, in fact, generalizes the availability of the same bis-retinoid fluorophores (in terms of their absorbance) irrespective of the genetic origin of mouse. Finally, we plot the distinct absorbance spectra (**Fig.2.d.**) of the three bis-retinoid fluorophores by spectral decomposition, applied to the average plot of **I.** & **II.** in **Fig. 2.c**. It is interesting to compare here the bis-retinoid fluorophores we obtained from our collective *in vivo* studies with the bis-retinoid fluorophores reported (ex-vivo) so far, in literature [17-20]. It is observed from **Fig.2.d.** that, the peak absorption of the short wave (SW: 437nm) is exactly matching with the absorption peak of the A2E (438nm) where, the mid wave (MW: 482nm) and long wave (LW: 520nm) agree well with A2-DHP-PE (490nm) and all-trans-retinal dimer-E (510nm), respectively. Here, we believe that the offset between our *in vivo*-results and *ex vivo*-reports lies within the acceptance range.

**Table 1.** Lists the estimated ratio of radiative to non-radiative decay time ($\tau_r/\tau_{nr}$) for short wave (SW), mid wave (MW) and long wave (LW) retinoid-fluorophores observed in Abca4 and C57BL/6 mice retina. The value of $\tilde{\nu}$, $\nu_0$, $\nu^{edge}$, and $\nu_0^{edge}$ are obtained directly from Fig.2.b. (see the supplementary for details).

| Abca4 | $\tilde{\nu}$ (nm) | $\nu_0$ (nm) | $\nu^{edge}$ (nm) | $\nu_0^{edge}$ (nm) | $(\tilde{\nu} - \nu^{edge})$ (nm) | $(\nu_0 - \nu_0^{edge})$ (nm) | $\tau_r/\tau_{nr}$ |
|---|---|---|---|---|---|---|---|
| SW | 529 | 430 | 535 | 450 | 6 | 20 | ~2.33 |
| MW | 603.6 | 470 | 656 | 490 | 52.4 | 20 | ~0.62 |
| LW | 649.4 | 510 | 660.5 | 550 | 11.1 | 40 | ~2.60 |
| C57BL/6 | | | | | | | |
| SW | 518.4 | 440 | 537.1 | 470 | 18.7 | 30 | ~0.60 |
| MW | 564.3 | 480 | 638.1 | 500 | 73.8 | 20 | ~0.73 |
| LW | 631.5 | 510 | 661.4 | 550 | 29.9 | 40 | ~0.34 |

**Bis-retinoid imaging:** Apart from identifying the bis-retinoid products based on their *in vivo* native interactions, REES can be used as an excellent probe for imaging. Particularly, the built-in fluorescence anisotropy is exploited as an imaging-probe in the context of REES [6,8,9]. Note that, in general, not all fluorescence-emissions are anisotropic, but it depends on the depolarizing events available in the system [6,8]. Here is the thumb rule that, the existence of REES in a system always ensures the availability of higher emission-anisotropy [6,9,16], having maximum value at FC- state (unrelaxed state) and minimum or ideally, zero at R- state (completely relaxed state) because of the depolarizing events of $\vartheta$ (**Fig.1.a. subset**), connecting the FC- and R- states. So, it is important to utilize the emission-anisotropy from near-FC state for each bis-retinoid fluorophore during their anisotropy imaging. **Fig. 3.a** shows the steady state fluorescence anisotropy spectra from three bis-retinoid fluorophores, excited at their peak- excitation wavelengths at 520nm, 480nm and 430nm for R3(: all-trans-retinal dimer-E), R2(: A2-DHP-PE) and R1(: A2E), respectively.

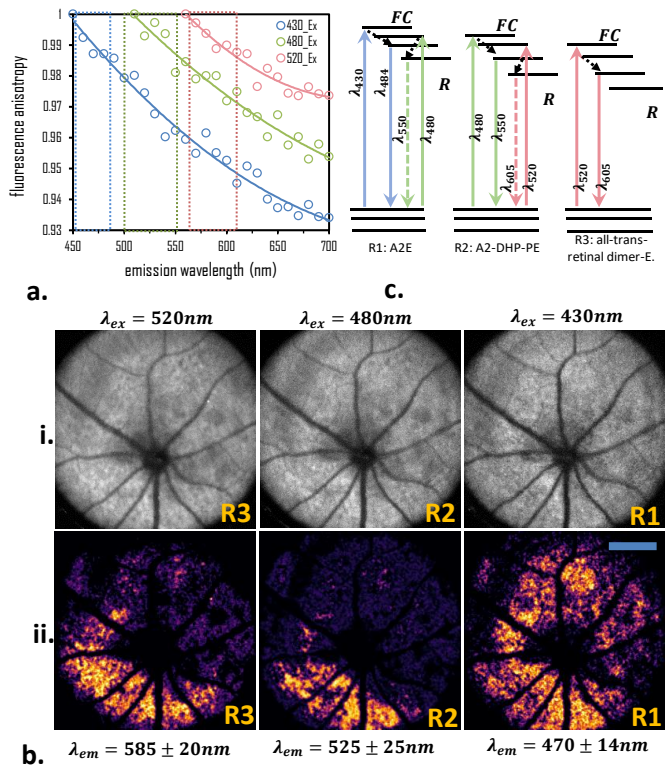

**Fig.3. Retinoid fluorophores imaging.** **a.** Steady state emission-anisotropy spectra. The rectangular boxes showing the emission bands chosen for imaging the retinoid products. **b.** row-**i.** Auto-fluorescence intensity images obtained through unpolarized excitations at 520nm, 480nm, 430nm and, unpolarized detections from 585±20nm, 525±25nm and 470±14nm emission bands. Row-**ii.** Anisotropy distributions of bis-retinoid products are obtained through selective polarized-excitations at 520nm, 480nm and, 430nm and collected two mutually orthogonal-polarized image components for each excitation from their respective emission bands (as above) to perform finally, image subtraction with ImageJ software; scale bar 250µm **c.** Potential energy diagram of REES (energy-labels are not in scale) for three bis-retinoid products. The number on the downward vertical lines represents the right-most edge of each emission band chosen for imaging.

A marked difference is observed between the intensity-based (row **i.**) and anisotropy-based (row **ii.**) images of the bis-retinoid (**Fig.3.b.**). It is observed that, for a selective excitation, the intensity distribution of its corresponding bis-retinoid product is not well-resolved in the intensity-based images. Because the background intensity from the other sub-excited bis-retinoid are overlapped. In contrast, anisotropy-based images can show anisotropy distribution of the respective bis-retinoid, exclusively. The reason can be understood from the potential energy diagram of REES for the three bis-retinoid fluorophore systems (R1, R2, & R3) as per our present case (**Fig.3.c.**). The selected excitations at 520nm, 480nm, and 430nm for imaging the respective bis-retinoid products are shown by upward continuous lines (**Fig.3.c.**). It is illustrated that, there is a possibility of cross-talking when we select, for example, 520nm for R3, it can sub-excite, at least, the nearest neighboring energy band of R2. However, for anisotropy imaging, the emission from R3 contributes more than that of R2. Because, the emission for R3 originates from near-FC-state whereas, the emission for R2 originates from near-R-state. As we discussed early that the emission from R- state is more isotropic in nature, which means the emission is almost equally available in the parallel (P) and perpendicular (S) state of the polarized-images. As a result, the emission component shown by the downward dotted line from near-R-state of R2 is likely to be nullified (or unselected) once we subtract the images (P & S) for anisotropy mapping but, the emission from near-FC of R3, shown by continuous downward line, is more likely to present (or selected) in the anisotropy images for 520nm excitation. Similarly, excitation at 480nm, and 430nm are likely to image exclusively, the anisotropy distribution of A2-DHP-PE and A2E, respectively.

**Conclusion:** We have developed and systematically demonstrated an *in vivo*, native interaction based spectroscopic probe for imaging the bis-retinoid products in mouse retina. We begin with observing the native interaction mediated typical spectral response as predicted in REES and analyze it to quantify the number of underlying fluorophores. The quantification adopts three approaches namely, identification of the number of Strokes-shifts, estimation of the typical ratio of radiative and nonradiative decay dynamic of the fluorophores, and measurement of *in vivo* action spectrum. Finally, we demonstrate the anisotropy distribution of the bis-retinoid products by exploiting the built-in anisotropy property of REES.

Beside this imaging probe, understanding of REES in bis-retinoid fluorophores can address a few shortcomings observed in other existing techniques available for imaging the bis-retinoid products. For example, life-time based imaging technique[21-23] finds some ambiguity in measuring the lifetime of yellow flecks[22] appeared in Stargardt disease affected eye. Here, the realization of REES in retinoid system conveys the message that the bis-retinoid fluorophores system is an anti-Kasha[16] system where the excited state life-time of these bis-retinoid fluorophores depends on the native interaction. This means the life-time of the yellow flecks can vary depending on their physical size, shape, viscosity and polarity of the native environment. Further, non-linear microscopy-based imaging techniques[12-14] finds a fundamental limitation that is the requirement of high excitation energy, something close to the damage level of the retina. In contrast, our anisotropy-based imaging technique is likely to map the bis-retinoid products exclusively, at regular power level. We believe that this native interaction based sophisticated probe can be adopted widely for early stage detection of Stargardt disease[21-23], age related macular degeneracy (AMD)[24-28] in retina. Moreover, the potential of differentiating these bis-retinoid fluorophores seemingly leads to a deeper understanding of the photochemical pathway for these diseases.

***Correspondence:** skmanna@ucdavis.edu

**Acknowledgement**
We thank Professor Edward Pugh, Jr. and Dr. Robert J. Zawadzki for useful discussions and providing living animal imaging facility in Eyepod. The studies were partially supported by EY026555 (NEI Core Grant).